\documentclass{aastex62}

\received{January 1, 2018}
\revised{January 7, 2018}
\accepted{\today}
\submitjournal{ApJ}

\shorttitle{No SMPs in WR\,2}
\shortauthors{Chen\'e et al.}
\begin{document}

\title{No detection of Strange Mode Pulsations in massive prime candidate}

\correspondingauthor{Andr\'e-Nicolas Chen\'e}
\email{andrenicolas.chene@gmail.com}

\author[0000-0002-1115-6559]{Andr\'e-Nicolas Chen\'e}
\affil{Gemini Observatory, Northern Operations Center, 670 North A'ohoku Place, Hilo, HI 96720, USA}

\author{Nicole St-Louis}
\affiliation{D\'epartement de Physique, Universit\'e de Montr\'eal, C. P. 6128, succ. centre-ville, Montr\'eal (Qc) H3C 3J7, and\\
 Centre de Recherche en Astrophysique du Qu\'ebec, Canada}

\author{Anthony~F.~J. Moffat}
\affiliation{D\'epartement de Physique, Universit\'e de Montr\'eal, C. P. 6128, succ. centre-ville, Montr\'eal (Qc) H3C 3J7, and\\
 Centre de Recherche en Astrophysique du Qu\'ebec, Canada}

\author{Olivier Schnurr}
\affiliation{Leibniz-Institut f\"ur Astrophysik Potsdam (AIP), An der Sternwarte 16, 14482 Potsdam, Germany}

\author{\'Etienne~Artigau}
\affiliation{D\'epartement de Physique, Universit\'e de Montr\'eal, C. P. 6128, succ. centre-ville, Montr\'eal (Qc) H3C 3J7, and\\
 Centre de Recherche en Astrophysique du Qu\'ebec, Canada}

%% Note that the \and command from previous versions of AASTeX is now
%% depreciated in this version as it is no longer necessary. AASTeX 
%% automatically takes care of all commas and "and"s between authors names.

%% AASTeX 6.2 has the new \collaboration and \nocollaboration commands to
%% provide the collaboration status of a group of authors. These commands 
%% can be used either before or after the list of corresponding authors. The
%% argument for \collaboration is the collaboration identifier. Authors are
%% encouraged to surround collaboration identifiers with ()s. The 
%% \nocollaboration command takes no argument and exists to indicate that
%% the nearby authors are not part of surrounding collaborations.

%% Mark off the abstract in the ``abstract'' environment. 
%\begin{abstract}
%
%Due to its compactness and high temperature, WR\,2 is one of the best candidates for strange mode pulsations (SMPs) with expected periods of 10 to 30\,minutes. However, our high precision photometric observations fails at revealing such pulsations.

%\end{abstract}

%% Keywords should appear after the \end{abstract} command. 
%% See the online documentation for the full list of available subject
%% keywords and the rules for their use.

%% From the front matter, we move on to the body of the paper.
%% Sections are demarcated by \section and \subsection, respectively.
%% Observe the use of the LaTeX \label
%% command after the \subsection to give a symbolic KEY to the
%% subsection for cross-referencing in a \ref command.
%% You can use LaTeX's \ref and \label commands to keep track of
%% cross-references to sections, equations, tables, and figures.
%% That way, if you change the order of any elements, LaTeX will
%% automatically renumber them.
%%
%% We recommend that authors also use the natbib \citep
%% and \citet commands to identify citations.  The citations are
%% tied to the reference list via symbolic KEYs. The KEY corresponds
%% to the KEY in the \bibitem in the reference list below. 

\section{Introduction} \label{sec:intro}

Theoretical work suggests that strange mode pulsations (SMPs) are present in hot and luminous stars with a high luminosity-to-mass ratio, where the thermal timescale is short compared to the dynamical timescale and radiation pressure dominates \citep[][]{Gl93}. The most violent SMPs are expected in classical WR stars, the bare, compact helium-burning cores of evolved massive stars \citep[][]{Gl99}, where SMPs are predicted to manifest themselves in cyclic photometric variability with periods ranging from minutes to hours and amplitudes in the several to tens of mmag range. Since WR\,2 is one of the hottest WR stars known, it should make it an excellent candidate for SMPs.

\section{Observations and reduction}\label{Phot}

We monitored WR\,2 using the NIR panoramic camera CPAPIR \citep{Ar04} at the 1.6m telescope of the Observatoire du Mont M\'egantic (QC, Canada) from 2009 January 11 to 16, securing 3 useful 3-6 hour continuous sequences (Fig.\,\ref{figPh}a). Each exposure was 10.8\,s long, with 5\,s overhead. The data were obtained in ``staring mode'', i.e. the telescope pointing was kept fixed on the sky through the entire sequence \citep{Gi13}, which renders our results fairly independent of flat-field errors. In order to isolate the light from the deepest layers of the WR wind, we selected a narrow-band filter ($\Delta\lambda = 0.025\,\rm{\mu m}$) centered on a region of the spectrum where no strong emission lines are present ($\lambda_c = 2.033\,\rm{\mu m}$).

All images were uniformly reduced using standard procedures carried out with routines written in Interactive Data Language (IDL). We performed aperture photometry using the {\it aper} Astrolib routine \citep{La93} on all stars present in the WR\,2 field. We adopted an aperture size equal to twice the FWHM of the point spread function (PSF) and an annulus of sky was selected with inner and outer radii of respectively 4 and 8 times the FWHM of the PSF. The light curves were obtained by correcting the zero-point of all frames, using the best frame obtained during the first 30\,mins of observation of each night. Since no flux standards are known in the band we used, we retain instrumental magnitudes (roughly calibrated to the K-band 2MASS photometry).

\section{Light curve analysis}

\begin{figure*}
\gridline{\fig{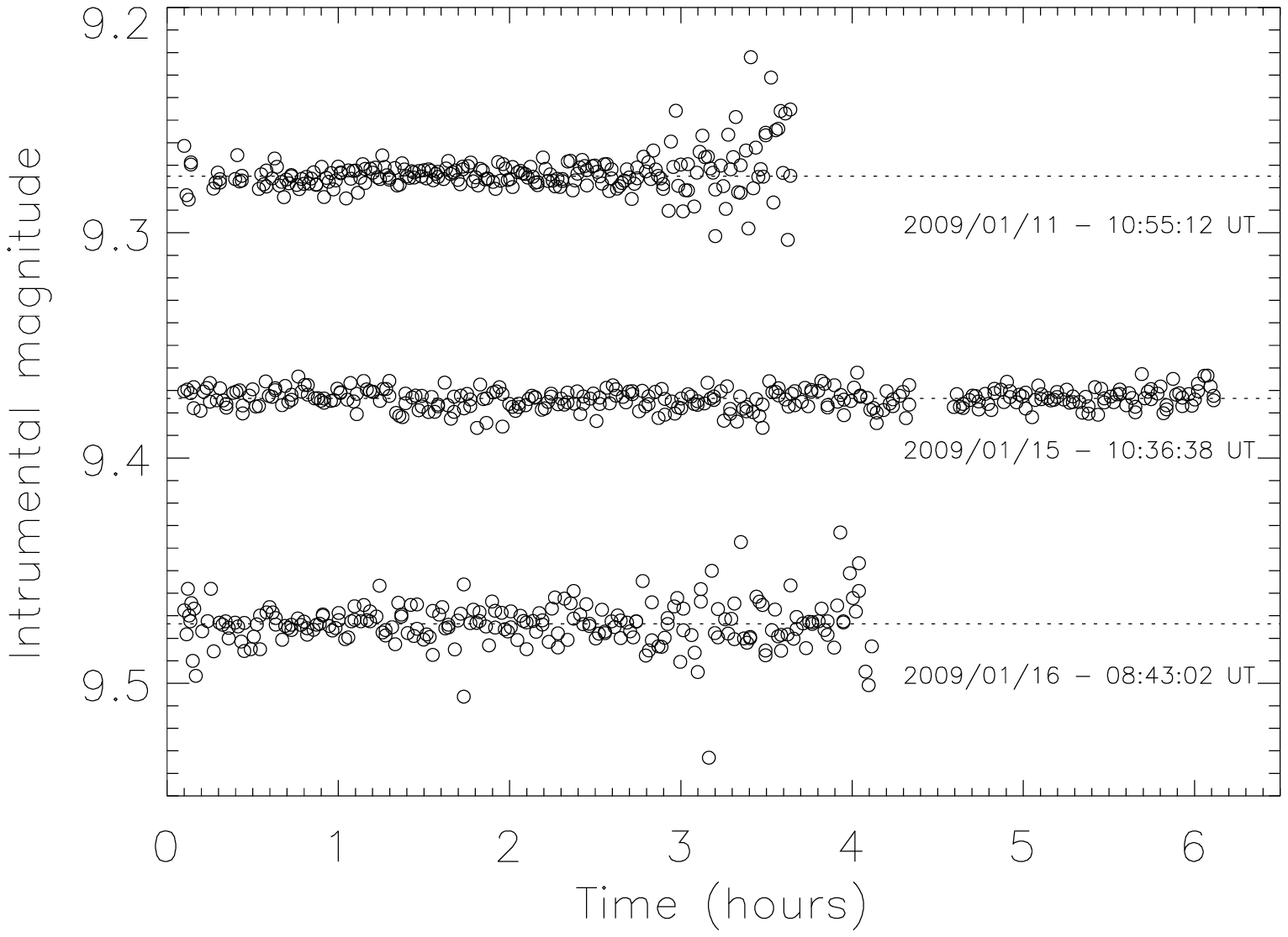}{0.5\textwidth}{(a)}
\fig{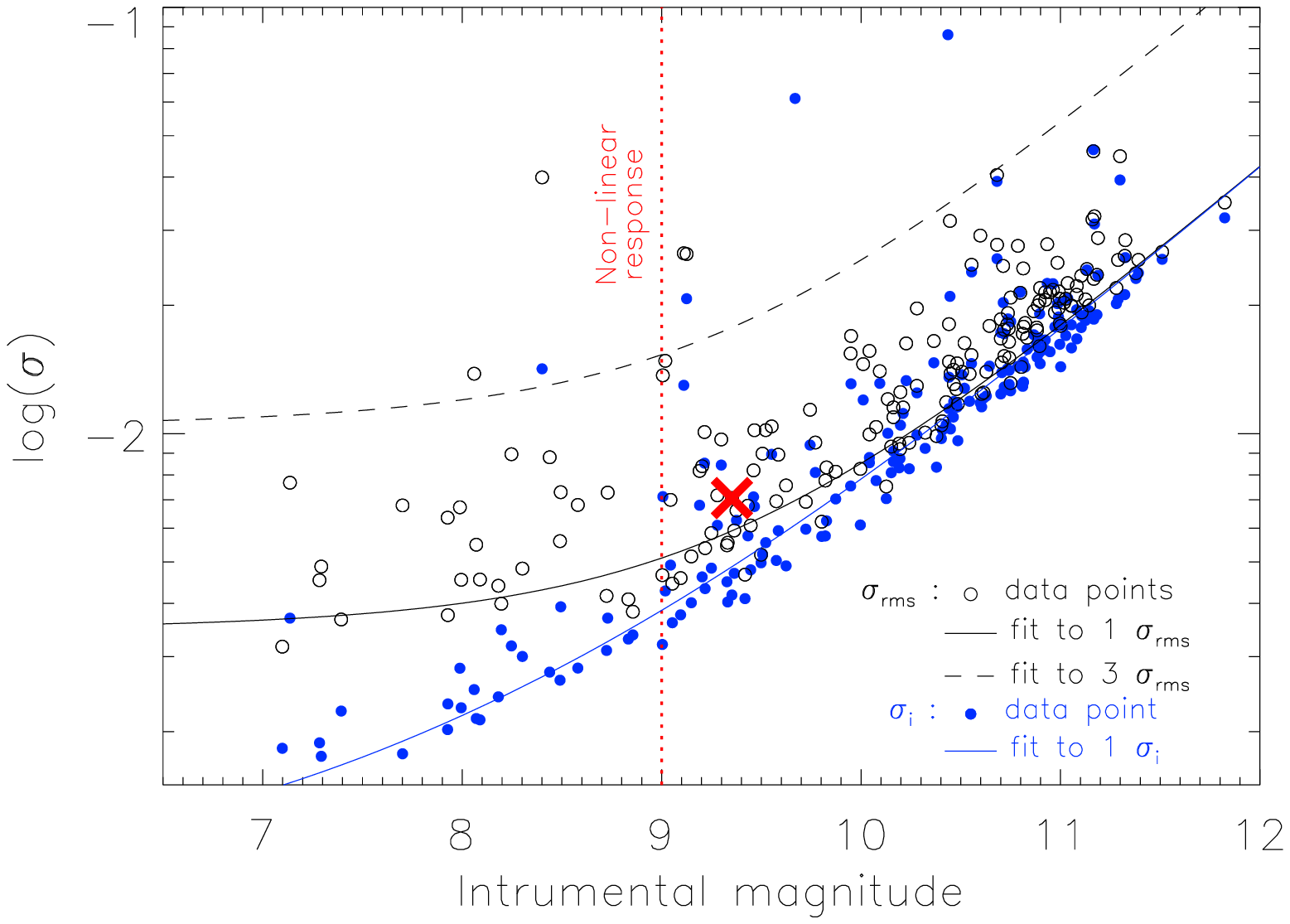}{0.5\textwidth}{(b)}
}
\gridline{\fig{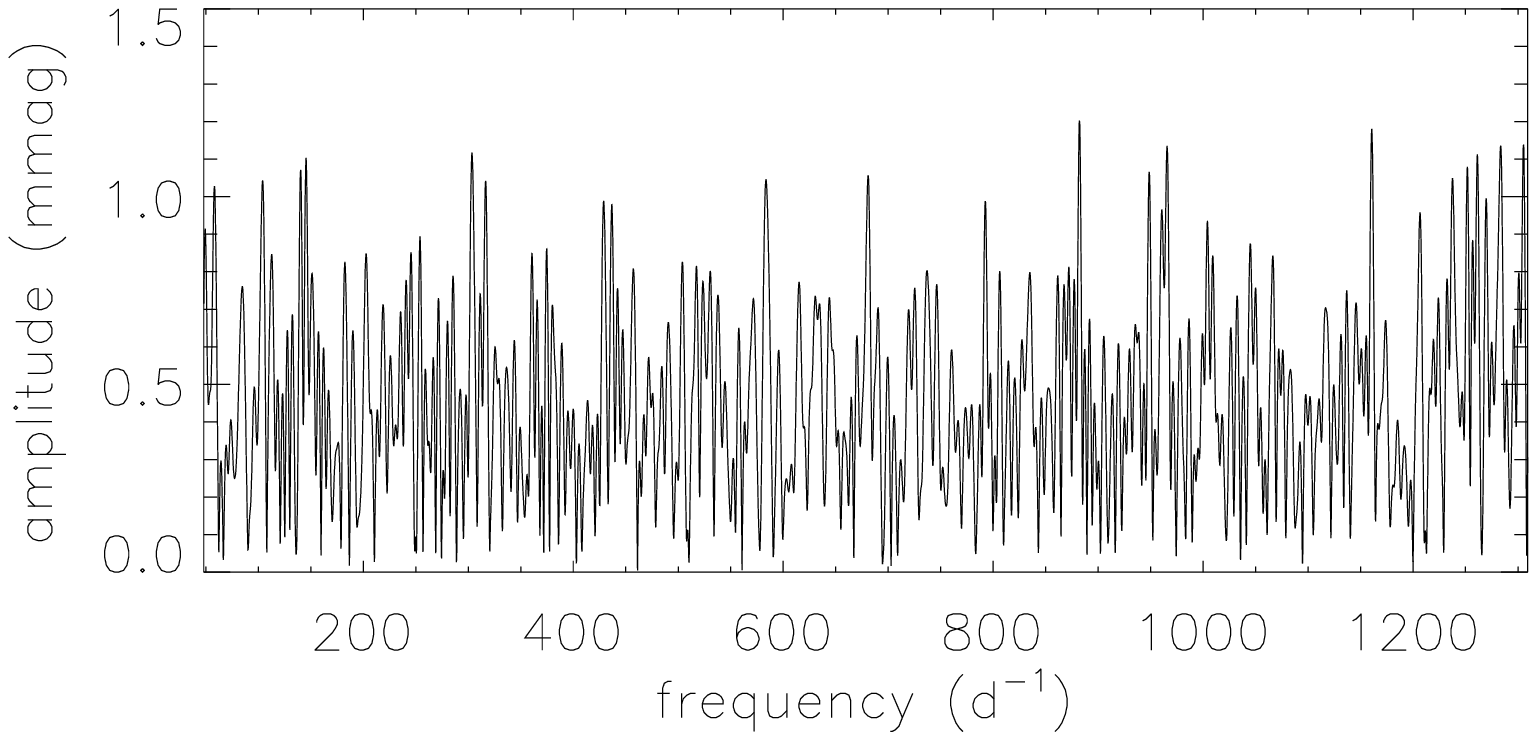}{0.6\textwidth}{(c)}
}
 \caption{{\it a)} CPAPIR light curves of WR\,2 for 2009 January 11, 15 and 16. The data are arbitrarily shifted from one night to the next. {\it b)} Standard deviations of the light curves of all detected stars (including WR\,2) during the night of 2009 January 15, as a function of the average instrumental magnitude using two different definitions \citep[$\sigma_{{\rm rms}}$ and $\sigma _{i}$, as defined in][]{De11}. 
%The first is the standard $\sigma_{{\rm rms}}$, plotted using open, black circles. The second one is $\sigma_i$ \citep[defined in][]{De11}, plotted using blue dots. The bottom blue and black solid lines are fits to the data ($\sigma _{i}$  and $\sigma_{{\rm rms}}$ respectively) of the theoretical expression between the magnitude and the photometric accuracy (derived using the principle of propagation of errors). 
The $\sigma_{{\rm rms}}$ value of WR\,2 is marked with a red X. 
%One vertical red dotted line at 9\,mag marks where the stars enter into a nonlinear regime. 
{\it c)} Periodogram of the light curve obtained in 2009 January 15.}
  \label{figPh}
\end{figure*}

In Fig.\,\ref{figPh}b, WR\,2 is much closer to the 1- than to the 3-$\sigma_{{\rm rms}}$ limit, and can be confidently labeled as non variable at that limit. Also, WR\,2's $\sigma_i$ gives a $R_\sigma$ value of 1.7 \citep[for complete definition, see][]{De11}, which confirms that WR\,2 does not vary significantly on timescales up to $\sim$6\,hours with an upper limit on the amplitude variation of 4\,mmag. This result is also confirmed during the two other observing nights.

Visual inspection of the WR\,2 light curve does not show any significant intrinsic change in time. Atmospheric variations cause a larger scatter at the end of 2 nights. We used a phase-dispersion minimization (PDM) algorithm \citep{St78} as well as a Scargle algorithm \citep{Sc82,Ro87} to search for periodicity in the photometric observations between periods of 1 minute and 0.5 day, using a step in frequency of 0.003\,d$^{-1}$. This analysis failed to reveal any stable period, and only led to a noisy periodogram with no significant peak (see Fig.\,\ref{figPh}c). The highest peak in the periodogram has an amplitude of 1.2\,mmag, which can be considered the smallest amplitude of a coherent periodic variation that we could have detected. 

\section{Conclusions}\label{Disc}

Taking the predictions from \citet{Gl99} for a star with $M=14M_\odot$, $\log{L/L_\odot}=5.41$ and $\log{T_{\rm{eff}}}=5.089$, i.e. within 10\% of the values obtained by \citet{Ha06} for WR\,2, we see that the expected SMPs should have an amplitude of 80\,mmag over a period of $\sim15$\,mins. This is more than one order of magnitude greater than our detection limit. Only the model for a low-mass WR star from \citet{Gl99} predicts an amplitude of 4\,mmag over a period of $\sim200$\,secs for SMPs or in other words, 1-$\sigma_{{\rm rms}}$ for our dataset. However, the stellar parameters (such as the temperature and luminosity) used for that model are both 50\% lower than those of WR\,2. Also, if we use $\sigma_i$ (=2.5\,mmag) as the formal error on each point and if we consider that during the night of 2009 January 15 we covered the predicted period $\sim100$ times with $\sim15$ data points per phase interval, we should expect some significant signal in the periodogram, or at least some coherent signal once the curve is folded into the predicted period. This is not what we observe.

%% If you wish to include an acknowledgments section in your paper,
%% separate it off from the body of the text using the \acknowledgments
%% command.
\acknowledgments

ANC is supported by the Gemini Observatory, which is operated by the Association of Universities for Research in Astronomy, Inc., on behalf of the international Gemini partnership of Argentina, Brazil, Canada, Chile, and the United States of America. AFJM and NSL acknowledge financial support from the Natural Sciences and Engineering Research Council (NSERC) of Canada. 

%% To help institutions obtain information on the effectiveness of their 
%% telescopes the AAS Journals has created a group of keywords for telescope 
%% facilities.
%
%% Following the acknowledgments section, use the following syntax and the
%% \facility{} or \facilities{} macros to list the keywords of facilities used 
%% in the research for the paper.  Each keyword is check against the master 
%% list during copy editing.  Individual instruments can be provided in 
%% parentheses, after the keyword, but they are not verified.

\vspace{5mm}
\facilities{OMM}

%% Similar to \facility{}, there is the optional \software command to allow 
%% authors a place to specify which programs were used during the creation of 
%% the manusscript. Authors should list each code and include either a
%% citation or url to the code inside ()s when available.

\software{Astrolib \citep{La93}  
          }


\begin{thebibliography}{}

\bibitem[\protect\citeauthoryear{Artigau et al.}{2004}]{Ar04} Artigau, \'E., Doyon, R., Vallee, P., Riopel, M. \& Nadeau, D. 2004, SPIE, 5492, 1479
\bibitem[\protect\citeauthoryear{Dekany et al.}{2011}] {De11} Dekany, I., M. Catelan, D. Minniti \& the VVV Collaboration 2011, arxiv.1111.0909
\bibitem[\protect\citeauthoryear{Girardin, Artigau \& Doyon}{2013}]{Gi13} Girardin, F., Artigau, \'E. \& Doyon, R. 2013, ApJ, 767, 61
\bibitem[\protect\citeauthoryear{Glatzel et al.}{1999}]{Gl99} Glatzel, W., Kiriakidis, M., Chernigovskij, S., \& Fricke, K.~J. 1999, MNRAS, 303, 116
\bibitem[\protect\citeauthoryear{Glatzel \& Kiriakidis}{1993}]{Gl93} Glatzel, W. \& Kiriakidis, M. 1993, MNRAS, 262, 85 
\bibitem[\protect\citeauthoryear{Hamann, Gr\"afener \& Liermann}{2006}]{Ha06} Hamann, W.-R., Gr\"afener, G. \& Liermann, A. 2006, A\&A, 457, 1015
\bibitem[\protect\citeauthoryear{Landsman}{1993}]{La93} Landsman, W. B 1993 in Astronomical Data Analysis Software and Systems II, A.S.P. Conference Series, Vol. 52, ed. R.~J. Hanisch, R.~J.~V. Brissenden \& Jeannette Barnes,  p. 246.
\bibitem[\protect\citeauthoryear{Roberts}{1987}]{Ro87} Roberts, D. H., Lehar, J., \& Dreher, J. 1987, AJ, 93, 968
\bibitem[\protect\citeauthoryear{Scargle}{1982}]{Sc82} Scargle, J. D. 1982, ApJ, 263, 835
\bibitem[\protect\citeauthoryear{Stellingwerf}{1978}]{St78} Stellingwerf, R. F. 1978, ApJ, 224, 953

\end{thebibliography}
\end{document}